\documentstyle[12pt,epsf,psfig]{article}
\begin{document}
\title{\bf Constraints on the Mass and Mixing of the 4th Generation Quark
From Direct CP Violation $\epsilon^{\prime}/\epsilon$ and Rare $K$ Decays }
\author{ Chao-Shang Huang${}^{a}$,~~~Wu-Jun Huo${}^{a,b}$, ~~~and Yue-Liang Wu${}^{a}$ \\ {\sl
${}^{a}$Institute of Theoretical Physics, Academia Sinica, P.O. Box $2735$},\\{\sl
 Beijing $100080$, P.R.  China}\\ {\sl
${}^{b}$Institute of High Energy Physics, Academia Sinica, P.O. Box $918(4)$},\\{\sl
 Beijing $100039$, P.R.  China}}

\date{}
\maketitle
\begin{abstract}

 We investigate the $\epsilon^{\prime} /\epsilon$ for $K\rightarrow \pi\pi$
 in a sequential fourth generation model. By giving the basic formulae
 for $\epsilon^{\prime}/\epsilon$ in this  model, we analyze the  numerical 
 results which are dependent of $m_{t^{\prime}}$ and imaginary part of the 
 fourth CKM factor,  $\mbox{Im}V^{*}_{t^{'}s}V_{t^{'}d}$ (or 
 $V^{*}_{t^{'}s}V_{t^{'}d}$ and the fourth generation CKM matrix phase $\theta$).
 We find that, unlike the SM, when taking the central values of all parameters 
 for $\epsilon^{\prime}/\epsilon$, the values of $\epsilon^{\prime}/ \epsilon$ 
 can easily fit to the current experimental data for all values of hadronic 
 matrix elements estimated from various approaches. Also, we show that the 
 experimental values of $\epsilon^{\prime}/\epsilon$ and rare K decays can 
 provide a strong constraint on both mass and mixing of the fourth generation 
 quark. When taking the values of hadronic matrix elements from the lattice or 
 $1/N$ expansion calculations, a large region of the up-type quark mass 
 $m_{t^{\prime}}$ is excluded.
\end{abstract}

 \vspace*{0.5cm} \noindent
PACS: 11.30.Er,12.60.-i,13.25.-Es,14.80.-j

\newpage

\section{Introduction}

Although the Standard Model (SM) is  very successful for explaining the 
particle physics experiments, it has to face the difficulties of many 
interesting open questions, such asCP violation. The new experimental results 
for $\epsilon^{\prime}/ \epsilon$, which measures direct CP violation in 
$K\rightarrow \pi\pi$ decays, have been reported by KTeV collaboration at 
Fermilab\cite{KTeV} and NA48 collaboration at CERN\cite{NA48},
\begin{eqnarray}
\mbox{Re}(\epsilon^{\prime}/\epsilon)& = & (28.0\pm 4.1) 
\cdot 10^{-4} \qquad {\rm KTeV},  \\
\mbox{Re}(\epsilon^{\prime}/\epsilon)& = & (18.5\pm 7.3) 
\cdot 10^{-3} \qquad {\rm NA48},
\end{eqnarray}
while the new world average reads\cite{NA48,ave}:
\begin{eqnarray}
\mbox{Re}(\epsilon^{\prime}/\epsilon) &=&  (21.1\pm 4.6) \cdot 10^{-4}
\end{eqnarray}
This establishment of direct CP violation rules out old superweak 
models\cite{swm}. Yet while the SM predicts a non-vanishing 
$\epsilon^{\prime}/ \epsilon$, the values in (1), (2) and (3) exceed
most theoretical predictions of SM\cite{bur,burr}. People have to face and 
resolve this discrepancy. Some possibilities to accommodate the data in SM 
have been pointed out~\cite{sber,hkps}.

The SM makes precise assumptions on the mechanism that generates the CP
 violation. The only source of CP violating phase originates from the 
 elements $V_{u_{i}d_{j}}$ of the CKM matrix with three quark generations. 
 In SM, there are both indirect ($\epsilon$) and direct ($\epsilon^{\prime}$) 
 CP violation.The analysis of $\epsilon^{\prime}/ \epsilon$ can be divided 
 into the short-distance (perturbative) part and long-distance 
 ( non-perturbative)part. Using the effective Hamiltonian, 
(${\cal H}_{W}= \sum_i C_i (\mu)Q_i (\mu)$\cite{buras,Ham}),
 one can obtain an expression of $\epsilon^{\prime}/\epsilon$
that involves CKM parameters ($V_{u_{i}d_{j}}$), Wilson coefficients ($y_i$)
and local operator matrix elements ($\langle Q_i\rangle_I$).
 The source of most theoretical uncertainties for 
 $\epsilon^{\prime}/ \epsilon$ is mainly from the difficulty in calculating
 non-perturbative part (local operator matrix elements), 
 comparing with the phenomenological determination of CKM
 parameters \cite{data} and the calculation of the Wilson coefficients at a 
 NLO leval\cite{Ham}. For $\epsilon^{\prime}/\epsilon$, one of the goals of 
 SM is to determine the hadronic matrix elements\cite{hadr,had1,gupt,had2,
 had3,bi}. 
 
 The interesting in this note is not in this non-perturbative part
 but the new effects with the fourth sequential generation particles in 
 the short-distance part.Except for the SM explanation, there are many 
 directions in the search for New Physics beyond the SM \cite{new,Sup,ferm,
 scal,gau,vec,ster} to resolve CP violation. Unlike SM, almost any extension 
 of SM has,in general, new CP violating phases. That is to say, they give 
 new CP violation sources. The new physics on CP violation beyond SM includes 
 CP violation in Supersymmetry models\cite{Sup}
and extensions of fermion sector\cite{ferm,vec,ster},
 scalar sector\cite{scal} and gauge sector\cite{gau} of SM.
In extensions of fermion sector, there are many models,
such as vector-like quark models\cite{vec}, sterile neutrino models\cite{ster},
proposed for probing new effects on CP violation.

In this note, like in ref.\cite{cwx}, we consider
 a sequential fourth generation model\cite{cwx,Mck}, in which an up-type quark 
 $t^{'}$, adown-type quark $b^{'}$, a lepton $\tau^{'}$, and a heavy neutrino 
 $\nu^{'}$ are added into the SM. The properties of these new fermions are 
 all the same as their corresponding counterparts of other three generations 
 except their masses and CKM mixing, see tab.1,
 
\begin{table}[htb] 
\begin{center} 
\begin{tabular}{|c||c|c|c|c|c|c|c|c|} 
\hline 
& up-like quark & down-like quark & charged lepton &neutral lepton \\ 
\hline 
\hline 
& $u$ & $d$& $e$ & $\nu_{e}$ \\ 
SM fermions& $c$&$s$&$\mu$&$\nu_{\mu}$ \\ 
& $t$&$b$&$\tau$&$\nu_{\tau}$\\
\hline
\hline new
fermions& $t^{'}$&$b^{'}$&$\tau^{'}$&$\nu_{\tau^{'}}$ \\ 
\hline 
\end{tabular}
\end{center}
\caption{The elementary particle spectrum of SM4}
\end{table}

  As the SM does not fix the number of generations,
 so far we don't know why there is more than one generation and what 
 law of Nature determines their number. On the one hand, the purely sequential
4th generation is constrained, even excluded in many literatures\cite{huo1}.
For example, in refs. \cite{willey,truong} the method of Pad\'{e} approximates is used to
show that for a large fermion mass, it is possible to dynamically generate
$p$-wave resonance and then the S parameter bound can serve to exclude  a heavy
fourth generation of fermions\cite{truong}.
Ref. \cite{beane} found that there is no violation of the S parameter upper
 bound for any value of the heavy fermion mass and that elastic unitarity,
 imposed as a constraint on strong $W_L W_L$ scattering, yields no information
 concerning and sheds no light on the existence of a heavy fourth generation.
 The ref. \cite{marciano} compared various precision determinations of the Femi
 constant $G_{F}$ to get the rather stringent bound of 3rd and 4th generation
 lepton mixing angle $\theta_{34}$. It founds  that the fourth charged lepton
 is too heavy and seems non-existence.
The precision electroweak measurements can also give
the strong  constrains to the sequential 4th generation, in particlar the
S parameter excludes it to 99.8\% CL if is degenerate, and if not and a small
T parameter is allowed then it is excluded to 98.2\% CL \cite{modif}.

However, on the other hand, experimentally, the LEP determinations
 of the invisible partial decay width of the $Z^0$ gauge boson only show that 
 there are certainly three {\it light} neutrinos of the usual type with mass 
 less than $M_{Z}/2$ \cite{Mark}. But the existence of the fourth generation
 with a heavy neutrino, i.e., $m_{\nu_4} \geq M_{Z}/2$ \cite{Berez} is not yet 
 excluded. Perhaps there exists some more deep or mechanism to give the room of
the sequential fourth generation. Because we really don't know why there only
three generations. So, it is not invalueble to research these new generation
as one of the new physics.
Before having a more fundamental reason for three generations,
 one may investigate phenomenologically whether the existing experimental data
  allow the existence of the fourth generation. This is also the main perpuse
of this note. There are a number of
  papers\cite{Mck,huo1} for discussing the fourth generation phenomena.
  
In our previous paper\cite{huo}, we have investigated the constraints on 
the fourth generation from the inclusive decays of $B\rightarrow X_sl^+l^-$ 
and $B\rightarrow X_s \gamma$. In this note, we further study its effects 
on direct CP-violating parameter $\epsilon^{\prime}/ \epsilon$ in 
$K\rightarrow \pi\pi$ decays as well as possible new constraints from
$\epsilon'/\epsilon$ and rare K decays.We limit ourselves to the non-SUSY
 case in order to concentrate on the phenomenological 
 implication of the fourth generation and will call this model 
 as SM4 hereafter for thesake of simplicity.
 
 CP-violating parameter $\epsilon^{\prime}/ \epsilon$ is a short distance 
 dominated process and is sensitive to new physics. In SM4 model, there 
 are not new operators produced. The new particle involved is only the fourth
  generation up-type quark$t^{\prime}$. The heavy mass of $t^{\prime}$ 
  propagating in the loop diagrams of penguin and box enters the Wilson 
  coefficients $y_{i}$, as well as top quark $t$ and $W$ boson. The effects
   of the fourth generation particles can only modify  $y_{i}$. Each new 
   Wilson coefficients$y^{\rm new}_i(\mu)$ is the sum of $y^{\rm sm}_i(\mu)$ 
   and $y^{(\rm 4)}_i(\mu)$ contributed by t and $t^{\prime}$ correspondingly.
    We can get  $y^{(\rm 4)}_{t^{\prime}}$ by taking the mass of $t^{\prime}$
    as one of the input parameter. Moreover, for obtaining
    $\epsilon^{\prime}/ \epsilon$ in SM4,we must know something about 
    elements $V_{{t^{\prime}}d_{j}}$ of the fourth generation $4\times 4$ 
    CKM matrix which now contains nineparemeters, i.e., six angles and three
     phases.But there are no any direct experimental measurements of them.
 So we have to get their information  indirectly from some meson decays.
  We investigate three rare $K$ decays, $K^{+}\rightarrow \pi^{+}\nu\bar\nu$,
  $K_{L}\rightarrow \pi^{0}\nu\bar\nu$ and  
  $K_{L}\rightarrow \mu^{+}\mu^{-}$\cite{kk},
  in SM4. These decays can give the constraint of the fourth CKM factor,
  $\mbox{Im}V^{*}_{t^{'}s}V_{t^{'}d}$
   (or $V^{*}_{t^{'}s}V_{t^{'}d}$ and a fourth generation phase $\theta$),
   which is need for calculating $(\epsilon^{\prime}/ \epsilon)^{\rm new}$.
   We shall take it as an additional input parameter. As a consequence,
the total $(\epsilon^{\prime}/ \epsilon)$ is the sum of
 $(\epsilon^{\prime}/ \epsilon)^{\rm sm}$ and
 $(\epsilon^{\prime}/ \epsilon)^{\rm 4}$contributed by the SM and the new 
 particle $t^{\prime}$ correspondingly. Unlike the SM, when taking the central 
 values of all parameters for $\epsilon^{\prime}/\epsilon$, the new value of
 $(\epsilon^{\prime}/ \epsilon)$ can reach the range of the current 
 experimental results whatever values of the non-perturbertion part, 
 hadronic matrix elements, are taken in all known cases. Also, the 
 experimental values of $(\epsilon^{\prime}/\epsilon)^{\rm exp}$
 impose strong constraints on the parameter space of 
 $\mbox{Im}V^{\ast}_{t^{\prime}s}V_{t^{\prime}d}$and $m_{t^{\prime}}$.
  
In sec. 2, we give the basic formulae for $\epsilon^{\prime}/ \epsilon$
with the fourth up-like quark $t^{\prime}$ in SM4.
In sec. 3, we analyze the constraints on the fourth generation CKM matrix
factor $\mbox{Im}V^{*}_{t^{'}s}V_{t^{'}d}$ which is necessary for 
calculating$\epsilon^{\prime}/ \epsilon$ in SM4. Sec. 4 is devoted to the 
numerical analysis. Finally, in sec. 5, we give our conclusion. 
  
\section{ Basic formulae for $\epsilon^{\prime}/\epsilon$ and Wilson 
Coefficients $y^{(\rm 4)}_i(\mu)$ in SM4 }

The essential theoretical tool for the calculation of 
$\epsilon^{\prime}/\epsilon$ is the $\Delta S=1$ effective 
Hamiltonian\cite{buras,Ham},
 \begin{equation}
 {\cal H}_{W}= \sum_i  \frac{G_F}{\sqrt{2}}  V_{ud}\,V^*_{us}
 \Bigl[z_i(\mu) + \tau\ y_i(\mu) \Bigr]  \; Q_i (\mu) \,
\end{equation}
with $\tau={V^{*}_{ts}V_{td}}/({V^{*}_{us}V_{ud}})$.
The direct CP violation in $K\rightarrow \pi\pi$ is described by $\epsilon^{\prime}$.
The parameter $\epsilon^{\prime}$ is given in terms of the  amplitudes
$A_0\equiv A(k\rightarrow(\pi\pi)_{I=0})$ and $A_2\equiv A(k\rightarrow(\pi\pi)_{I=2})$ 
as follows
\begin{equation}
\epsilon^{\prime}=-\frac{1}{\sqrt{2}}\xi(1-\Omega)\exp(i\Phi),  
\end{equation}
where 
\begin{equation}
 \xi=\frac{\mbox{Im}A_0}{\mbox{Re}A_0},\,\omega=\frac{\mbox{Re}A_2}{\mbox{Re}A_0},\,
 \Omega=\frac{1}{\omega}\frac{\mbox{Im}A_2}{\mbox{Im}A_0}
\end{equation}
and $\Phi=\pi/2+\delta_2-\delta_0\approx\pi/4$. With the effective Hamiltonian (4),
we can cast (5) into the form
\begin{equation}
\frac{\epsilon'}{\epsilon} = 
\mbox{Im}\lambda_t\cdot \left[ P^{(1/2)} - P^{(3/2)} \right],
\end{equation}
where
\begin{eqnarray}
P^{(1/2)} &=& \sum P_i^{(1/2)} = r \sum y_i \langle Q_i\rangle_0
(1-\Omega_{\eta+\eta'}) \\
P^{(3/2)} &=& \sum P_i^{(3/2)} = \frac{r}{\omega}
\sum y_i \langle Q_i\rangle_2.
\end{eqnarray}
with $r=G_F \omega/(2|\mbox{Re}A_0)|)$. 
$y_{i}$ are the Wilson coefficients and
the hadronic matrix elements are
\begin{equation}
\langle Q_i\rangle_I \equiv \langle (\pi\pi)_I | Q_i | K \rangle
\end{equation}
The operators $Q_i$ and $\langle Q_i\rangle_I$ are 
given explicitly in many reviews\cite{buras,Ham}

 When including the contributions from the fourth generation up-type quark
$t^{\prime}$, the above equations will be modified.
The corresponding effective Hamiltonian can be expressed as
 \begin{equation}
 {\cal H}_{W}= \sum_i  \frac{G_F}{\sqrt{2}}  V_{ud}\,V^*_{us}
 \Bigl[z_i(\mu) + \tau\ y^{\rm SM}_i(\mu)+
 \tau^{\prime}y^{(\rm 4)}_i(\mu) \Bigr]  \; Q_i (\mu) \, .
\end{equation}
with $\tau = {V^{*}_{ts}V_{td}}$ and $\tau^{\prime}={V^{*}_{t^{'}s}V_{t^{'}d}}$. 
In comparison with the SM, one may introduce the new effective
coefficient functions $y^{\rm new}_i(\mu)$
\begin{equation}
y^{\rm new}_{i}(\mu)=y^{\rm SM}_i+\frac{\tau'}{\tau} \cdot 
y^{(\rm 4)}_i(\mu),
\end{equation}
where $y_i(\mu)$ are the Wilson coefficient functions in the SM and $y_i^{(4)}$ are the ones
due to fourth generation quark contributions. The evolution for $ y^{(\rm 4)}_i(\mu)$ 
is analogy to the one $y^{\rm SM}_i(\mu)$ in SM\cite{buras,Ham}
except replacing the t-quark by  $t^{\prime}$ quark. The corresponding
diagrams of penguin and box are shown in fig. 1. 


Using (11) and (12), eq. (7) can be written as
\begin{eqnarray}
(\frac{\epsilon^{\prime}}{\epsilon})&=& (\frac{\epsilon'}{\epsilon})^{\rm SM}+
(\frac{\epsilon^{\prime}}{\epsilon})^{(4)}, \nonumber  \\
(\frac{\epsilon^{\prime}}{\epsilon})^{(4)}&=&
\mbox{Im}\lambda_t^{\prime}\cdot \left[ P^{\prime(1/2)} - P^{\prime(3/2)} \right],
\end{eqnarray}
where the definitions of $P^{\prime(1/2)}$ and $P^{\prime(3/2)}$ are the same as (8) and (9) only by 
changing $y_i(\mu)$ into $y^{(\rm 4)}_i(\mu)$, and
\begin{equation}
\mbox{Im}\lambda_{t^{\prime}}=\mbox{Im}V^{*}_{t^{'}s}V_{t^{'}d}.
\end{equation}

Thus the main test of evaluating $\epsilon'/\epsilon$ in the SM4 is to calculate 
the Wilson coefficients $y^{(\rm 4)}_i(\mu)$ and to provide the possible constraints on
$\mbox{Im}\lambda_{t^{\prime}}$. The constraints of 
$\mbox{Im}\lambda_{t^{\prime}}$ will be discussed
in next section. The calculation of $y^{(\rm 4)}_i(\mu)$ is the same as 
 their counterpart $y^{\rm SM}_i(\mu)$ in SM and can be simply done 
by changing $m_t$ to $m_{t^{'}}$,
which is easy to be found in any corresponding 
reviews\cite{buras,Ham}. Here we repeat the same calculations and only provide the numerical 
results for $y^{(\rm 4)}_i(\mu)$ as the functions of the mass $m_t^{'}$. 
In the numerical calculations we  take a large range for 
$t^{'}$-quark mass $m_{t'}=$ 50GeV, 100GeV, 150GeV, 200GeV, 250GeV, 300GeV, 400Gev
\cite{Mck} See tab. 2,

\begin{table}[htb]
\begin{center}
\begin{tabular}{|c||c|c|c|c|c|c|c|c|}
\hline
$m_t^{'}$(GeV) &50    &100    & 150 &200   & 250  &300   &350   &400     \\
\hline
$z_1^{\rm 4}$           &-0.594&-0.594&-0.594&-0.594&-0.594&-0.594&-0.594&-0.594  \\
$z_2^{\rm 4}$           & 0.323& 0.323& 0.323& 0.323& 0.323& 0.323& 0.323& 0.323  \\
\hline
\hline
$y_3^{\rm 4}$           & 0.028& 0.032& 0.036& 0.042& 0.048& 0.055& 0.064& 0.074  \\
$y_4^{\rm 4}$           &-0.049&-0.052&-0.056&-0.059&-0.064&-0.069&-0.075&-0.081  \\
$y_5^{\rm 4}$           & 0.011& 0.011& 0.012& 0.012& 0.013& 0.013& 0.014& 0.014  \\
$y_6^{\rm 4}$           &-0.089&-0.097&-0.112&-0.104&-0.107&-0.111&-0.114&-0.118  \\
\hline
\hline
$y_7^{\rm 4}/\alpha$    &-0.114&-0.076&-0.004& 0.092& 0.210& 0.348& 0.506& 0.686  \\
$y_8^{\rm 4}/\alpha$    &-0.034& 0.011& 0.097& 0.210& 0.350& 0.514& 0.704& 0.917  \\
$y_9^{\rm 4}/\alpha$    &-0.367&-0.825&-1.335&-1.913&-2.571&-3.318&-4.159&-5.098  \\
$y_{10}^{\rm 4}/\alpha$ & 0.172& 0.397&0.6475& 0.932& 1.255& 1.622& 2.037& 2.498  \\
\hline
\end{tabular}
\end{center}
\caption[]{$\Delta S=1 $ Wilson coefficients at $\mu=1.0{\rm GeV}$ for
 $\Lambda^{(\rm 4)}=340{\rm MeV}$ and $f=3$ effective flavors at
 leading order.
$y_1^{\rm 4} = y_2^{\rm 4} \equiv 0$.}
\end{table}

\section{Constraints on CKM Factor $V^{*}_{t^{'}s}V_{t^{'}d}$ in SM4}

Though we have no direct information for the additional fourth generation CKM matrix elements,  
while constraints may be obtained from some rare meson decays. In ref.\cite{huo}, we obtained the values
of the fourth CKM factor $V^{*}_{t^{'}s}V_{t^{'}b}$ from the decay of
 $B\rightarrow s\gamma$. In this paper, we shall investigate three
rare $K$ meson decays: two semi-leptonic decays $K^{+}\rightarrow \pi^{+}\nu\bar\nu$
and $K_{L}\rightarrow \pi^{0}\nu\bar\nu$, and one leptonic decay
$K_{L}\rightarrow \mu^{+}\mu^{-}$\cite{kk} within SM4. 
These decays can provide certain constraints on the fourth generation CKM factors, 
$V^{*}_{t^{'}s}V_{t^{'}d}$ , $\mbox{Im}V^{*}_{t^{'}s}V_{t^{'}d}$ and 
$\mbox{Re}V^{*}_{t^{'}s}V_{t^{'}d}$ respectively. 

Within SM, the decays $K\rightarrow\pi\nu\bar\nu$ 
 are loop-induced semileptonic FCNC process determined
only by $Z^0$-penguin and box diagram.  These decays are the theoretically
cleanest decays in rare K-decays. 
The great virtue of $K_L\rightarrow\pi^0\nu\bar\nu$ is that it proceeds
almost exclusively through direct CP violation \cite{lib} which is 
very important for the investigation of $\epsilon^{'}/\epsilon$ in SM4.
The precise calculation of these two decays 
at the NLO in SM can be found in Refs\cite{kk1}. While experimentally,
its branching ratio has not yet been well measured, only an upper bound has be given and is 
larger by one order of magnitude than the one  
in SM (see tab. 3)\footnote{From ref.\cite{ygr}, one can easily derive by means of
isospin symmetry the following model independent bound:
\begin{eqnarray*}
 Br(K^0\to\pi^+\nu\bar{\nu}) < 4.4 \cdot Br(K_L\to\pi^+\nu\bar{\nu})
 \end{eqnarray*}
 which gives
 \begin{eqnarray*}
  Br(K^0\to\pi^+\nu\bar{\nu}) < 6.1\times 10^{-9}
  \end{eqnarray*}  This bond is much stronger than the direct experimental bound.}. 
 This remains allowing the New Physics to dominate their decay amplitude\cite{new}. Moreover,  Unlike
the previous two semi-leptonic decays, the branching ratio $Br(K_L\rightarrow\mu^+\mu^-)$ has already 
been measured with a very good precision. While its experimental result is several 
times larger than theoretical prediction in SM (see tab. 3).
This also provides a window for New Physics.

\begin{table}[htb]
\begin{center}
\begin{tabular}{|c||c|c|c|c|}
\hline
    & $Br(K^+\to\pi^+\nu\bar{\nu})$ &  $Br(K_L\to\pi^0\nu\bar{\nu})$ &$Br(K_L\rightarrow\mu^+\mu^-$)  \\
\hline
\hline
Experiment & $<2.4 \times 10^{-9}$\cite{BNL} & $<1.6\times 10^{-6}$\cite{Ada}
 &
 $(6.9\pm 0.4)\times 10^{-9}$\cite{BNL1}  \\
&$(4.2+9.7-3.5)\times10^{-10}$\cite{adl} & $<6.1\times 10^{-9}$\cite{ygr}&$(7.9\pm 0.7)\times 10^{-9}$\cite{KEK}\\
\hline
SM & $(8.2\pm 3.2)\times 10^{-11}$\cite{buc}& $(3.1\pm 1.3)\times 10^{-11}$ \cite{buc} &
 $(1.3\pm0.6)\times 10^{-9}$\cite{gab}  \\
\hline
\end{tabular}
\end{center}
\caption{ Comparison of $B(K^+\to\pi^+\nu\bar{\nu})$, $B(K_L\to\pi^0\nu\bar{\nu})$
and $B(K_L\to\pi^0\nu\bar{\nu})$ among  
the experimental values and SM predictions with maximum mixing.}
\end{table}

In the SM4, the branching ratios of the three decay modes mentioned 
above receive additional contributions from the up-type quark $t^{\prime}$ \cite{new1}
 \begin{equation}
Br(K^+\to\pi^+\nu\bar{\nu})=\kappa_+\left| \frac{V_{cd}V_{cs}^*}{\lambda}
P_0+\frac{V_{td}V_{ts}^*}{\lambda^5}\eta_tX_0(x_t)+\frac{V_{t'd}
V_{t's}^*}{\lambda^5}\eta_{t'}X_0(x_{t'})\right|^2, 
\end{equation}
 \begin{equation}
Br(K_L\to\pi^0\nu\bar{\nu})=\kappa_L\left| \frac{{\rm Im}V_{td}V_{ts}^*}
{\lambda^5}\eta_tX_0(x_t)+\frac{{\rm Im}V_{t'd}
V_{t's}^*}{\lambda^5}\eta_{t'}X_0(x_{t'})\right|^2, 
\end{equation} 
\begin{equation}
Br(K_L\to\mu\bar{\mu})_{\rm SD}=\kappa_{\mu} \left[ \frac{{\rm Re}
\left( V_{cd}V_{cs}^*\right) }{\lambda}P'_0+\frac{{\rm Re}\left( V_{td} 
V_{ts}^*\right) }{\lambda^5}Y_0(x_t)+\frac{{\rm Re}\left( V_{t'd}
V_{t's}^*\right) }{\lambda^5}Y_0(x_{t'})\right]^2.
\end{equation}
where $\kappa_+,\kappa_L,\kappa_{\mu}$,$X_0(x_t)$ , $X_0(x_{t'})$,
$Y_0(x_t)$ ,$Y_0(x_{t'})$,$P_0, P'_0$ may be found in Refs\cite{buras,Ham}. 
The QCD correction factors are taken to be $\eta_t =$ 0.985 and $\eta_{t'}=$ 1.0 \cite{new1}.

To solve the constrains of the 4th generation CKM matrix factors $V^{*}_{t^{'}s}V_{t^{'}d}$,
${\rm Im} V^{*}_{t^{'}s}V_{t^{'}d}$ and ${\rm Re} V^{*}_{t^{'}s}V_{t^{'}d}$, we must conculate
the Wilson coefficients  $X_0(x_{t'})$ and  $Y_0(x_{t'})$. They are the founctions of the mass
of the 4th generation top quark, $m_{t'}$. Here we give their numerical results according to several
values of $m_{t'}$, (see table 4)
 \begin{table}[htb]
\begin{center}
\begin{tabular}{ |c|| c| c| c |c |c| c |c |c |c| }
\hline
  $m_{t'}$(GeV)& 50  & 100 &150   &200  &250 &300   & 400  & 500 & 600
 \\ \hline \hline
 $X_0(x_{t'})$  &0.404  &0.873  &1.357 & 1.884& 2.474& 3.137 &4.703  & 6.615 & 8.887   \\
 \hline
\hline
  $Y_0(x_{t'})$  & 0.144  & 0.443 &0.833 & 1.303 &1.856&2.499   &4.027  & 5.919 & 8.179   \\
 \hline
\end{tabular}
\end{center}
\caption{Wilson coefficients $X_0(x_{t'})$, $Y_0(x_{t'})$ to $m_{t'}$   }
\end{table}
We found that the Wilson coefficients  $X_0(x_{t'})$ and  $Y_0(x_{t'})$
increase with the  $m_{t'}$. To get the largest constrain of the factors in eq.
(15), (16) and (17), we must use the little value of $m_{t'}$. Considering that
 the 4th generation particles must have the mass larger than $M_Z /2$ \cite{Mark},
we take $m_{t'}$ with 50 GeV to get our constrains of those three factors.

Then, from (15), (16) and (17), we arrive at the following constraints
\begin{equation}
 |V^{*}_{t^{'}s}V_{t^{'}d}| \leq 2\times10^{-4},
 \end{equation}
 \begin{equation}
|\mbox{Im}V^{*}_{t^{'}s}V_{t^{'}d}| \leq 1.2\times 10^{-4},
\end{equation}
\begin{equation} 
| \mbox{Re}V^{*}_{t^{'}s}V_{t^{'}d}| \leq 1.0\times 10^{-4}.
\end{equation}
For the numerical calculations, we will take $|\mbox{Im}V^{*}_{t^{'}s}V_{t^{'}d}| \leq 1.2\times 10^{-4}$.

 It is easy to check that the equation (18) obeys the
 CKM matrix unitarity constraint, which states that any pair of rows, or
any pair of columns, of the CKM matrix are orthogonal.\cite{data}.  The relevant one to those decay 
channels is 
\begin{equation} V_{us}^{*}V_{ud}+V_{cs}^{*}V_{cd}+V_{ts}^{*}V_{td}
+V_{t^{'}s}^{*}V_{t^{'}d}=0.
 \end{equation} 
Here we have taken the average values of the SM CKM matrix
elements  from Ref. \cite{data}.
Considering the fact that the data of CKM matrix is not yet very accurate, there still exists 
a sizable error for the sum of the first three terms. Using the value of
$V^{*}_{t^{'}s}V_{t^{'}d}$ obtained from eq. (18), the sum of the four terms in 
the left hand of (21) can still be close to $0$,  
 because the values of $V^{*}_{t^{'}s}V_{t^{'}d}$ are about $10^{-4}$
order, ten times smaller than the sum of the first three ones in the left of (21). 
Thus, the values of  $V^{*}_{t^{'}s}V_{t^{'}d}$ remain
 satisfying the CKM matrix unitarity constraints in SM4 within the present uncertainties.

\section{The Numerical Analysis}

In the calculation of  $\epsilon^{\prime}/\epsilon$, the main source of uncertainty 
are the hadronic matrix elements $\langle Q_i \rangle_I$. They depend generally on 
the renormalization scale $\mu$ and on the scheme used to renormlize the operators $Q_i$.
But the calculation of $\langle Q_i \rangle_I$ is much beyond the perturbative method.
They only can be tread by the non-perturbative methods, like lattice methods, 
$1/N$ expansion, chiral quark models and chiral effective langrangians, which is not
sufficient to obtain the high accuracy. 
We shall present the analysis on $t^{'}$-quark effects when considering 
the uncertainties of $\langle Q_i \rangle_I$ due to model-dependent calculations.
 
It is customary to express the matrix elements $\langle Q_i \rangle_I$ in terms of 
non-perturbative parameters $B^{\rm (1/2)}_i$ and $B^{\rm (3/2)}_i$ as follows:
\begin{equation}
\langle Q_i\rangle _0 \equiv B^{\rm (1/2)}_i \langle Q_i\rangle^{(\rm vac)}_0;
\langle Q_i\rangle _2 \equiv B^{\rm (3/2)}_i \langle Q_i\rangle^{(\rm vac)}_2.
\end{equation}
The full list of $\langle Q_i \rangle_I$ is given in ref.\cite{hadr}. 
We take the phenomenological values of $B_i$\cite{bi} (see tab.5)
except for $B^{(1/2)}_6$ and $B^{(3/2)}_8$
which are taken as input parameter with values calculated by 
three different non-perturbative methods.  
Other numerical input parameters are given in tab.6.
\begin{table}[htb]
\begin{center}
\begin{tabular}{ |c| c| c| c |c |c| c |c |c |c| }
\hline
  $B_1^{\rm(1/2)}$ & $B_2^{\rm(1/2)}$ & $B_3^{\rm(1/2)}$ & $B_4^{\rm(1/2)}$
 & $B_5^{\rm(1/2)}$ &${\bf B_6^{\bf(1/2)}}$&  $B_7^{\rm(1/2)}$ & $B_8^{\rm(1/2)}$
 & $B_9^{\rm(1/2)}$ & $B_{10}^{\rm(1/2)}$
 \\ \hline
 $13.0$ & $6.1\pm1.0$ & $1.0^{*}$ & $5.2^{*}$ & $B_6^{\rm(1/2)}$ &{\bf INPUT}& $1.0^{*}$
 & $1.0^{*}$ & $7.0^{*}$ & $7.5^{*}$   \\
 \hline
 \hline
  $B_1^{\rm (3/2)}$ & $B_2^{\rm(3/2)}$ & $B_3^{\rm(3/2)}$ & $B_4^{\rm(3/2)}$
 & $B_5^{\rm(3/2)}$ &  $B_6^{\rm(3/2)}$ &$B_7^{\rm(3/2)}$ & ${\bf B_8^{\bf(3/2)}}$&
  $B_9^{\rm(3/2)}$ & $B_{10}^{\rm(3/2)}$\\
  \hline
  $0.48$ & $0.48$ & $1.0^{*}$ & $5.2^{*}$ & $1.0\pm0.3^{*}$ &$1.0\pm0.3^{*}$
 & $1.0^{*}$ & {\bf INPUT} & $0.48$ & $0.48$\\
 \hline
\end{tabular}
\end{center}
\caption{Phenomennolegical values of $B_i$. $^{*}$ stands for an "educated guess".}
\end{table}

\begin{table}[htb]
\begin{center}
\begin{tabular}{ |c| c|| c| c ||c |c| }
\hline
$\mbox{Re}A_0$& $3.33\times 10^{-7}$GeV& $\Omega_{\eta\eta^{'}}$& $0.25$& 
 $G_{\rm F}$&$1.166\times10^{-5}$GeV${^{-2}}$\\
 $\mbox{Re}A_2$& $1.50\times 10^{-8}$GeV& $\omega$     & $0.045$ &$\mbox{Im}\lambda_t$& 
 $1.34\times10^{-4}$ \\
 \hline\hline
 $m_d(m_c)$ & $8  $MeV & $m_\pi$    & $138$MeV & $\Lambda^{\rm 4}_{\rm\overline{MS}}$ &3$40$MeV\\
 $m_s(m_c)$ & $130$GeV & $m_{\rm K}$& $498$MeV & $M_W$          & $80.2$GeV\\
 $m_c(m_c)$ & $1.3$GeV & $F_\pi$    & $131$MeV & $\alpha_s(M_Z)$& $0.117$  \\
 $m_b(m_b)$ & $4.8$GeV & $F_{\rm K}$& $160$MeV & $\alpha$       & $1/129$  \\
 $m_t(m_t)$ & $175$GeV &             &          & $\sin\theta_W$ & $0.23$   \\
 \hline
\end{tabular}
\end{center}
\caption{Neumerical values of the input parameters.}
\end{table}

We take the values of $B_6^{\rm(1/2)}$and $ B_8^{\rm(3/2)}$ in three non-perturbitive 
approaches, lattice methods, $1/N$ expansion and chiral quark models  (see
tab.7) and the figs (see figs.2, 3, 4) in each case respectively.
\begin{table}[htb]
\begin{center}
\begin{tabular}{ |c||c|c| c| }
\hline
&lattice method&  $1/N$ expansion & chiral quark models\\
\hline
\hline
  $ B_6^{\rm(1/2)}$ & $1.0\pm0.02\pm0.05$\cite{des} &$0.81$\cite{burr}&  $1.6\pm 0.3$\cite{sber} \\
   \hline
 $B_8^{\rm(3/2)}$  & $0.8\pm 0.15$\cite{gupt} & $0.49$\cite{burr} & $0.92\pm 0.002$\cite{sber} \\
\hline
\end{tabular}
\end{center}
\caption{The input values of $ B_6^{\rm(1/2)}$ and $B_8^{\rm(3/2)}$ in
three cases.}
\end{table}

The numerical results are shown in figs. 2,3,4 which correspond to the three cases of calculating hadronic matrix elements,
lattice method, 1/N eapansion, and chiral quark model, respectively.
We now present a study for $\epsilon^{'}/\epsilon$ as functions of $\mbox{Im}\lambda_{t^{\prime}}$ and 
$m_{t'}$: $\epsilon^{'}/\epsilon$ versus $\mbox{Im}\lambda_{t^{\prime}}$ with fixing
$m_{t^{'}}$ is plotted in figs. (a);
 $\epsilon^{'}/\epsilon$ versus $m_{t^{\prime}}$ with fixing  $\mbox{Im}\lambda_{t^{\prime}}$ 
is ploted in figs. (b); and the allowed parameter space of $\mbox{Im}\lambda_{t^{\prime}}$ 
and $m_{t^{'}}$  is plotted in figs. (c). We shall analyze each case in detail as follows.
 
 In figs (a) we plot eight lines corresponding to  $m_{t^{'}}=$50, 100, 150, 200 250, 300, 350, 400GeV 
respectively. First, we notice that the slope of the line
 decreases as $m_{t^{'}}$ increases. At a value of $m_{t^{'}}$, about 230GeV,
 the slope is  zero because the second part in the right hand side of eq. (13)  vanishes. 
 The reason is similar to that in SM, i.e., with increasing $m_{t^{'}}$ 
the EW penguins become increasingly important and
their contributions to $\epsilon^{'}/ \epsilon$ are with the opposite sign 
to those of QCD penguins so that at some values of 
$m_{t^{'}}$ there is a cancellation. The behavior comes essentially 
once $m_{t^{'}}$ becomes larger than 230GeV, the slope is negative. 
 Its absolute value increases with $m_{t^{'}}$. Such a behavior comes essentially   
 from the change of the Wilson coefficients $y_i^{(\rm 4)}$ as $m_{t^{'}}$.
 Second, from figs (a), we  found, within the constraints on $\mbox{Im}\lambda_{t^{\prime}}$ from
 the three rare $K$ meson decays, that $\epsilon^{'}/\epsilon$ can generally be consistent with the 
 experimental average except for some ranges of $m_{t^{'}}$ once the non-perturbative parameters
  $B_6^{\rm(1/2)}$ and  $B_8^{\rm(3/2)}$ are taken values calculated based on the lattice gauge theory 
  and $1/N$ expansion. Such a range roughly ranges from 170GeV to 300GeV, which 
can be seen from figs. (2a) and (3a). There is no excluded range for the case of the chiral quark model.
 This is because in the first two
  cases, the SM values $(\epsilon^{'}/\epsilon)^{\rm SM}$ are about $8.8\times10^{-4}$,
  which is much lower than the experimental average. For a large range of $m_{t^{'}}$, 
$(\epsilon^{'}/ \epsilon)^{(4)}$ is not large
enough to make total $\epsilon^{'}/\epsilon$ reach the experimental average. 
But in the chiral quark model,
  the SM value is about $18.8\times10^{-4}$ which is in the $1\sigma$ error range 
  of the present experimental average so that $\epsilon^{'}/\epsilon$ can 
reach the experimental average for all values of $m_{t^{'}}$ in the reasonable region. 
Thus once the non-perturbative method calculations become more reliable and the experimental
  measurements get more accuracy, it may provide more strong constraints 
  on the forth generation quark from the study on $\epsilon^{'}/\epsilon$. Unfortunately, we can't get 
  any information on the upper bound of $m_{t^{'}}$.
  
  We also plot in Figs (b) eight curves corresponding to $\mbox{Im}\lambda_{t^{\prime}}=
 $1.0, 0.75, 0.5, 0.25, -0.25, -0.5, -0.75, -1.0$\times10^{-4}$ respectively. 
Thus similar results as those in the figs (a) are arrived.
 These curves are divided into two types determined by the sign of 
 the fourth generation CKM factor $\mbox{Im}\lambda_{t^{\prime}}$. The reason is also similar to 
the analysis for figs. (a). The figs.(b) also show the constraints on $\mbox{Im}\lambda_{t^{\prime}}$.
 It is interesting to see that there is an excluded
region from 0 to $0.6\times10^{-4}$ based on lattice gauge theory results and 
from 0 to $0.76\times10^{-4}$ based on the $1/N$ expansion results. While there 
 is no such an excluded region based on the chiral quark model results. 
The reason is the same as that in the analysis of 
figs (a). Moreover, it seems that $\mbox{Im}\lambda_{t^{\prime}}$ favors the negative values which 
may be interesting since the negative value of $\mbox{Im}\lambda_{t^{\prime}}$
  is better to satisfy the unitarity constraints of the CKM matrix (see eq. (21)). 
  Therefore if there could exist the fourth generation, from both the theoretical and the experimental
  parts, one might be able get usefull information on the fourth generation CKM matrix elements,
  such as $V^{*}_{t^{'}s}V_{t^{'}b}$ which has been studied in our previous paper\cite{huo}. 

  In figs. (c), we show the correlation  between
 $\mbox{Im}\lambda_{t^{\prime}}$ and $m_{t^{'}}$. The three curves in the figure
 correspond to the experimental values of the new world average and 
 its $1\sigma$ error, respectively. It is seen that the the allowed parameter space is 
strongly limited for all three cases when the ratio $\epsilon^{'}/\epsilon$ is around
 the present experimental average within 1 $\sigma$ error. The 
allowed parameter space is divided into two pieces except in the chiral quark model. This is in agreement 
with  the analyses in figs. (a) and (b). Such a small parameter space indicates 
that $\epsilon^{'}/\epsilon$ may impose a very strong constraint on the mass and mixing of 
the fourth generation up-type quark.
 

\section{Conclusion}

 In summary, we have investigated the direct CP-violating parameter
 $\epsilon^{\prime}/\epsilon$ in $K^{0}-\bar K^{0}$ system with considering
the up-type quark $t^{\prime}$ in SM4. The basic formulae 
 for $\epsilon^{\prime}/\epsilon$ in SM4 has been presented and the 
Wilson coefficient functions in the SM4 have also been evaluated. The
numerical results of the additional Wilson coefficient functions
have been given as functions of the mass $m_{t'}$.
 We have also studied the relevant rare $K$ meson decays:
 two semi-leptonic decays, $K^{+}\rightarrow \pi^{+}\nu\bar\nu$
and $K_{L}\rightarrow \pi^{0}\nu\bar\nu$, and one leptionic decay 
$K_{L}\rightarrow \mu^{+}\mu^{-}$, which allow us to obtain the bounds on
the fourth generation CKM matrix factor  $V^{*}_{t^{'}s}V_{t^{'}d}$.
In particular, we have analyzed the numerical result of $\epsilon^{\prime}/\epsilon$ 
 as the function of $m_{t^{\prime}}$ and imaginary part of the fourth CKM factor, 
$\mbox{Im}V^{*}_{t^{'}s}V_{t^{'}d}$
 (or $V^{*}_{t^{'}s}V_{t^{'}d}$ and a fourth generation CKM matrix phase $\theta$).
 The correlation between $\epsilon^{\prime}/\epsilon$ and 
 $\mbox{Im}V^{*}_{t^{'}s}V_{t^{'}d}$ has been studied in detail with different 
hadronic matrix elements calculated from various approaches, such as
 lattice gauge method, $1/N$ expansion and chiral quark model. 
 It has been seen that, unlike the SM, when taking the central values of all 
 parameters, the values of  $\epsilon^{\prime}/ \epsilon$ can be easily made to be consistent with 
the current experimental data for all estimated values of 
the relevant hadronic matrix elements from various approaches.
Especially, we have also investigated the allowed parameter space of $m_{t^{'}}$
 and $\mbox{Im}V^{*}_{t^{'}s}V_{t^{'}d}$, as a consequence, when considering $1\sigma$-error 
 of the current experimental data for $\epsilon^{\prime}/\epsilon$,
 the allowed parameter space for  $m_{t^{'}}$
 and $\mbox{Im}V^{*}_{t^{'}s}V_{t^{'}d}$ is very small and strongly restricted. This implies
 that the experimental data in K system  can provide strong constraints on the mass of $t^{\prime}$-quark
 and also on the fourth generation quark mixing matrix.
 
\section*{Acknowledgments}
This research was supported in part by the National Science Foundation of China.

\newpage

\begin{figure}
\psfig{file=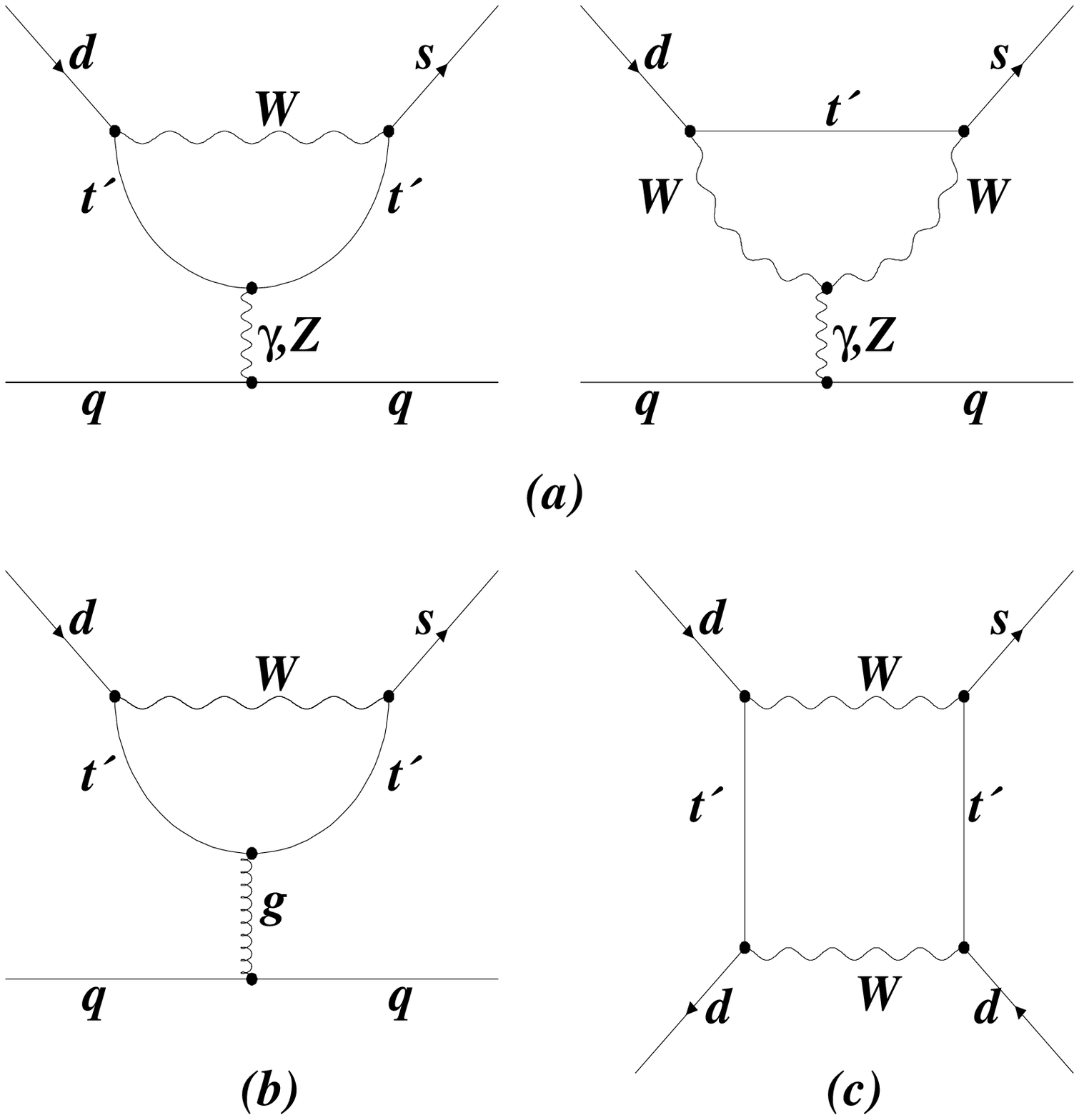,height=15.0cm}
\vskip .0cm

\caption{The Additional Diagrams of (a) EW, (b) QCD
 Penguins and (b) Box with $t^{'}$.}
\end{figure}

\begin{figure}
\vskip -3.0cm
\psfig{file=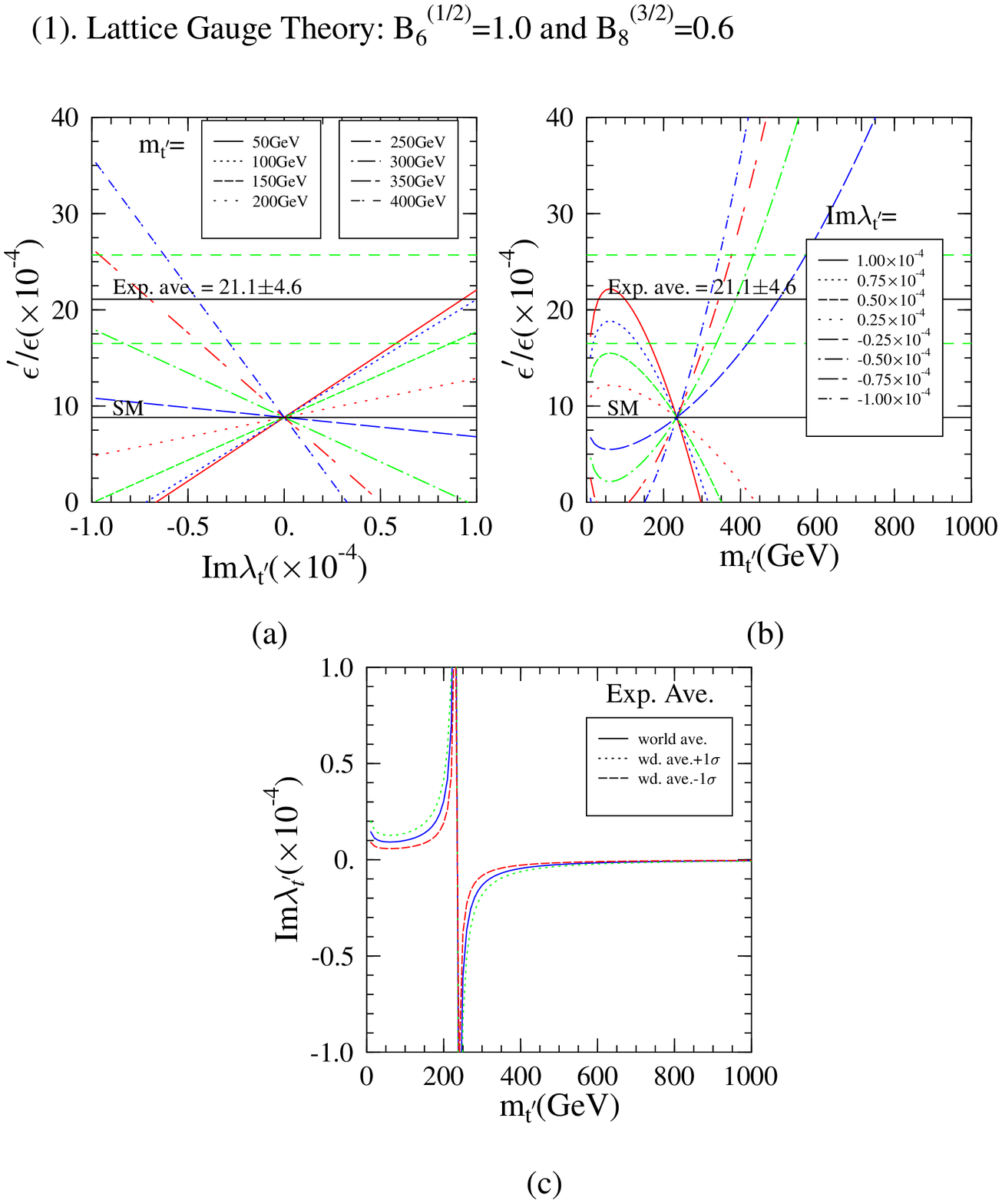,height=25.0cm}
\vskip -3.0cm
\caption{The Diagrams of (a) $\epsilon^{'}/\epsilon$ to  $\mbox{Im}\lambda_{t^{'}}$,
 (b)$\epsilon^{'}/\epsilon$ to $m_{t^{'}}$ and (c) 
 parameter space: $\mbox{Im}\lambda_{t^{'}}$ to $m_{t^{'}}$ with 
 $B_6^{\rm (1/2)}$ and $B_8^{\rm (3/2)}$ in Lattice Gauge Theory.}
\end{figure}

\begin{figure}
\vskip -3.0cm
\psfig{file=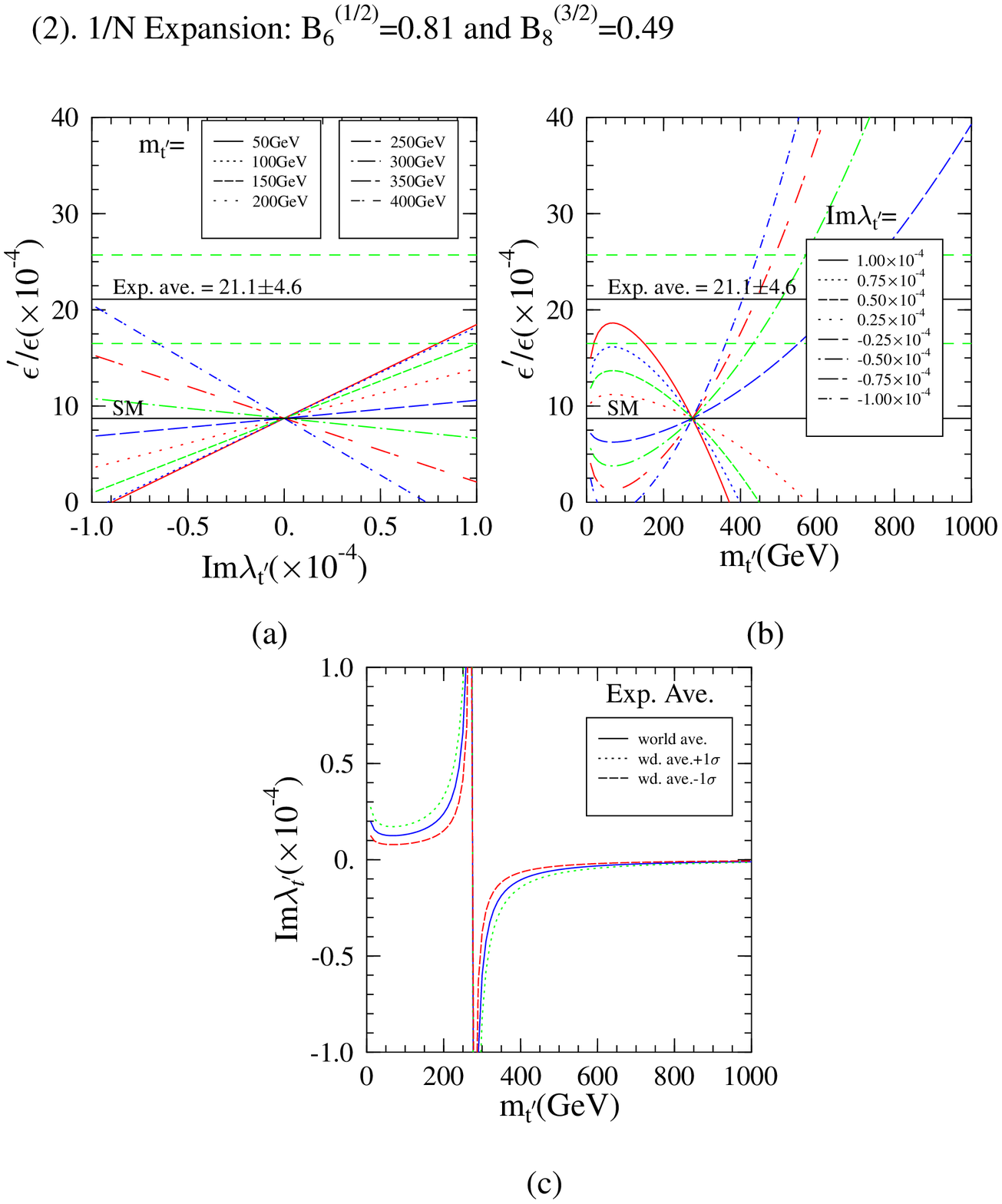,height=25.0cm}
\vskip -3.0cm
\caption{Same as fig. 1 in $1/N$ Expansion.}
\end{figure}

\begin{figure}
\vskip -3.0cm
\psfig{file=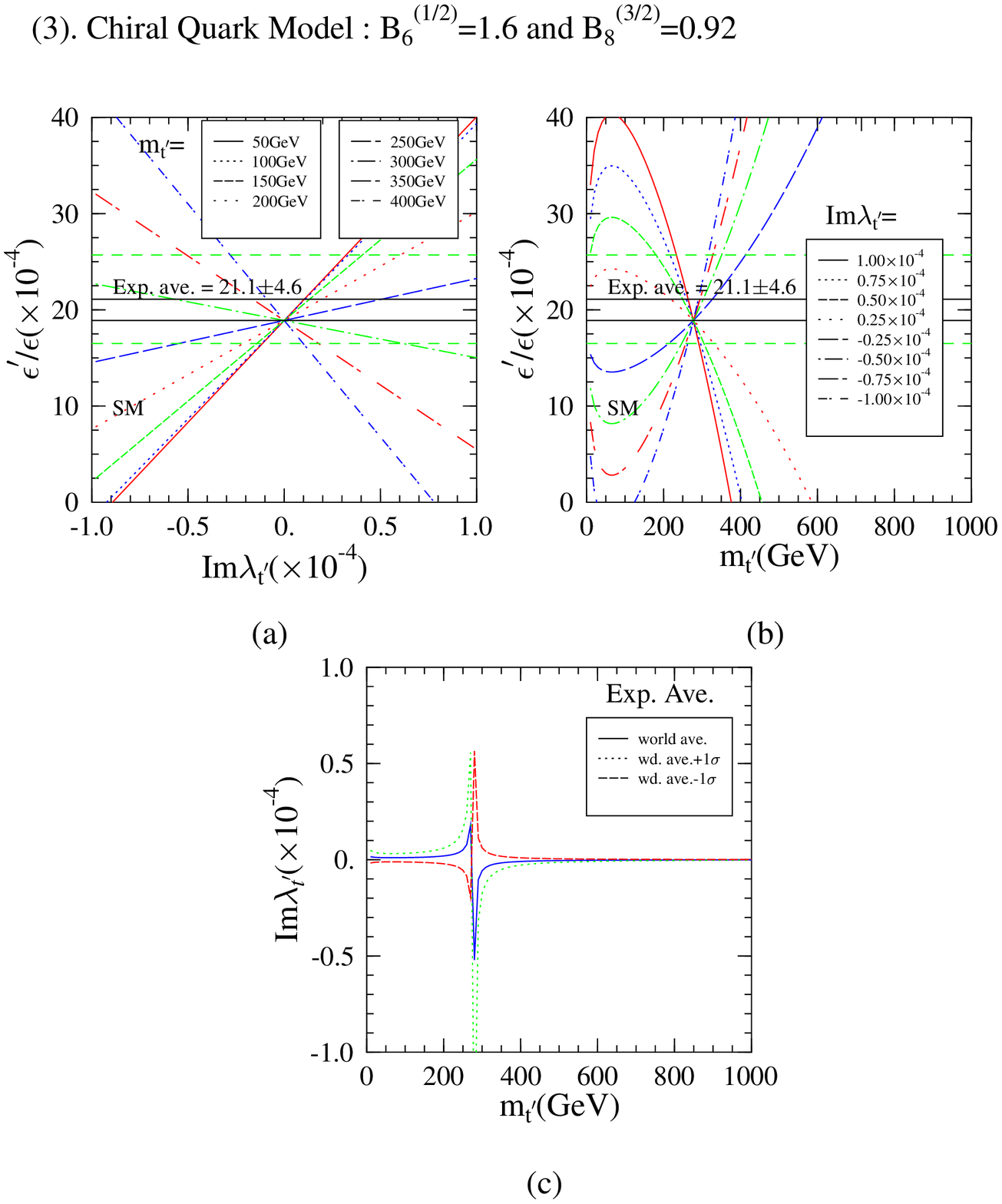,height=25.0cm}
\vskip -3.0cm
\caption{Same as fig. 1 in Chiral Quark Model.}
\end{figure}


\begin{thebibliography}{99}
\bibitem{KTeV} A. Alavi-Harati {\it et al.}, Phys. Rev. Lett. {\bf 83} (1999) 22.

\bibitem{NA48} V. Fanti {\it et al.}, Phys. Lett. {\bf 465} (1999) 335.

\bibitem{ave} U. Nierste, hep-ph/9910257; V. Fanti {\it et al},

\bibitem{swm}L. Wolfenstein, Phys. Rev. Lett. {\bf 13} (1964) 562.

\bibitem{bur}M. Fabbrichesi; hep-ph/9909224, hep-ph/0002235;
 M. Jamin, hep-ph/9911390.

\bibitem{burr}S. Bosch, A.J. Buras, M. GORBAHN, S. J\"ager, M. Jamin, M.E. Lauteubacher
and L. Silvestrni, hep-ph/9904408.

\bibitem{sber}S. Bertolini, J.O. Eeg, M. Fabbrichesi, hep-ph/9802405.

\bibitem{hkps}Y.-Y. Keum, U. Nierste, and A.I. Sanda, hep-ph/9903230;  T. Hambye, G.O. K$\ddot{o}$hler, E.A. Paschos,
and P.H. Soldan, hep-ph/9906434; S. Gardner and G. Valencia, hep-ph/9909202.

\bibitem{buras}A.J. Buras, hep-ph/9806471.

\bibitem{Ham} G. Buchalla, A.J, Buras, M.E. Lautenbacher, Rev. of Mod. Phys.
{\bf 68} (1996) 1125 and references therein; 
 E. A. Paschos, Y.L. Wu, Mod. Phys. Lett. {\bf A6} (1991) 93. 

\bibitem{data}C.Caso et al., (Particle Data Group), Eur. Phys. J. C{\bf 3} (1998) 1.

\bibitem{hadr} G. Buchalla, A.J, Buras, M.E. Lautenbacher, Nucl. Phys. {\bf 408} (1993) 209.

\bibitem{had1}R. Gupta, hep-ph/9801412;
   G. Kilcup, R.Gupta and S.R. Sharpe, Phys. Rev. {\bf D57} (1998) 1654;
   L. Conti, {\it et al}. hep-ph/9711053;
   T. Blum {\it et al.}, hep-lat/9908025.
\bibitem{gupt}R. Gupta, T. Bhattacharaya and S.R. Sharpe, Phys. Rev. {\bf D55} (1997) 4036.   

\bibitem{had2}W.A. Bardeen, A.J. Buras and J.-M. Gerard, Phys. Lett. {\bf B180} (1986) 133;
Nucl. Phys. {\bf B293} (1987) 787; Phys. Lett. {\bf B192} (1987) 138;
J. Heinrich, E.A. Paschos, J.-M. Schwarz, and Y.L. Wu, Phys. Lett. {\bf B279} (1992) 140;
Y.L. Wu, Int.J. Mod. Phys. {\bf A7} (1992) 2863;  
T. Hambye, G.O. K\"ohler, E.A. Paschos, P.H. Soldan and W.A. Bardeen, Phys. Rev. {\bf D58} (1998)
014017; W.A. Bardeen, Nucl. Phys. Proc. Suppl. {\bf 7A} (1989) 149.

\bibitem{had3}D. Espriu, E. de Rafael and J. Taron, Nucl. Phys. {\bf B345} (1990) 22;
J. Bijnens, Phys. Rept. {\bf 265} (1996) 369; 
S. Bertolini, J.O. Eeg, M. Fabbrichesi, Nucl. Phys. {\bf B449} (1995) 197;
Nucl. Phys. {\bf B476} (1996) 225; hep-ph/9802405;  
S. Bertolini, J.O. Eeg, M. Fabbrichesi and E.I. Lashin, Nucl. Phys. {\bf B514} (1998) 93.

\bibitem{bi}M. Faggrichesi, hep-ph/9909224.

\bibitem{new}  Y. Grossman, Y. Nir and R. Rattazzi, Hep-ph/9701231.

\bibitem{Sup}A.J. Buras, {\it et al.},hep-ph/9908371; A.L. Kagan and M.Neubert, hep-ph/9908404;
A.Masiero and H. Murayama, hep-ph/9903363; Chao-Shang Huang and Wei Liao, hep-ph/9908246, hep-ph/0001174 ;
M. Brhlik et al., hep-ph/9909480;  S. Baek and P. Ko, hep-ph/9909433; D.A. Demir, A. Masiero, O. Vives, hep-ph/9909325.

\bibitem{ferm}Y. Nir and H.R. Quinn, Ann. Rev. Nucl. Part. Sci. {\bf 42} (1992) 211;
Phys. Rev. {\bf D42} (1990) 1473; Y. Nir and D. Silverman, Nucl. Phys. {\bf B345} (1990) 301;
I. Dunietz, Ann. Phys. {\bf 184} (1988) 350; C.O. Dib, D.london and Y. Nir, Int. J. Mod. Phys. 
{\bf A6} (1991)1253; J.P. Silva and L. Wolfenstein, hep-ph/9610208;
Y. Grossman and M.P. Worah, Phys. Lett. {\bf B395} (1997) 241.

\bibitem{scal} Y.L. Wu and L. Wolfenstein, Phys. Rev. lett. {\bf 73}, 1762 (1994);
  L. Wolfenstein and Y.L. Wu , Phys. Rev. lett. {\bf 73}, 2809 (1994); 
  S. Weinberg, Phys. Rev. {\bf D42} (1990) 860; L. Lavoura, Int. J. Mod. Phys. {\bf A8} (1993)375;
Chao-Shang Huang and Shou-Hua Zhou, Phys. Rev. {\bf D61} (2000) 015011

\bibitem{gau} D. Chang, Nucl.Phys. {\bf B214} (1983) 435;
 H.Harari and M. Leurer, Nucl. Phys. {\bf B233} (1984) 221;
 M. Leurer, Nucl. phys. {\bf B226} (1986) 147; X.G. He, hep-ph/9903242.

\bibitem{vec} Y. Nir and D. Silverman, Phys. Rev. {\bf D42} (1990) 1477;
  W-S, Choong and D. Silverman, Phys. Rev. {\bf D49} (1994) 2322; L.T. Handoko, Hep-ph/9708447.

\bibitem{ster}V. Barger, Y.B. Dai, K. Whisnant and B.L. Young, Hep-ph/9901380;
R.N. Mohapatra, hep-ph/9702229; S. Mohanty, D.P. Roy and U. Sarkar, hep-ph/9810309;
S.C. Gibbons,{\it et al}., Phys. lett. {\bf B430} (1998) 296;
V. Barger, K. Whisnant and T.J. Weiler, Phys. lett. {\bf B427}, (1998) 97;
V. Barger, S. Pakvasa, T.J. Weiler and K. Whisnant, Phys. Rev. {\bf D58} (1998) 093016.

\bibitem{cwx}  K.C. Chou,  Y.L. Wu, and Y.B. Xie,  Chinese Phys. Lett. {\bf 1} (1984) 2.

\bibitem{Mck}J.F. Gunion, Douglas W. McKay, H. Pois, Phys.  Lett.  B{\bf 334} (1994) 339;
 Phys.  Rev.  D{\bf 51} (1995) 201.

\bibitem{huo1}references therein of Ref.\cite{huo}.

\bibitem{willey}R.S. Willey, Phys. Rev. {\bf D44} (1991) 3646;
H. Veltman and M. Veltman, Acta Phys. Pol. {\bf B22} (1991) 669.

\bibitem{truong}T.N. Truong, Phys. Rev. Lett. {\bf 70} (1993) 888.

\bibitem{beane}S.R. Beane and S. Varma, hep-ph/9304233.

\bibitem{marciano}W.J. Marciano, Phys. Rev. {\bf D60} (1999) 093006

\bibitem{modif}J. Erler and P. Langacker, http://www-pdg.lbl.gov/1999/stanmodelrpp.ps.

\bibitem{Mark}Mark II Collab., G.S. Abrams et al., Phys. Rev. Lett.
{\bf 63}(1989) 2173;
L3 Collab., B. Advera et al., Phys. Lett. B{\bf 231} (1989) 509;
OPAL Collab., I. Decamp et al., ibid., {\bf 231} (1989) 519;
DELPHI Collab., M.Z. Akrawy et al., ibid., {\bf  231} (1989) 539.

\bibitem{Berez}Z. Berezhiani and E, Nardi, Phys. Rev. D{\bf 52} (1995) 3087;
C.T. Hill, E.A. Paschos, Phys. Lett. B{\bf 241} (1990) 96.

\bibitem{huo} C.S. Huang, W. J. Huo and Y.L. Wu, Mod. Phys. Lett. {\bf A14} (1999)2453.

\bibitem{kk}R. D. Peccei, hep-ph/9909236;
 T. Hattori, T. Hasuike and S. Wakaizumi, hep-ph/9808412; A.J. Buras, hep-ph/9901409.

\bibitem{BNL}S. Adler, {\it et al}., Phys. Rev. Lett. {\bf B76} (1996) 1421.

\bibitem{Ada}J. Adams,{it et al},. hep-ex/9806007.

\bibitem{BNL1}A.P. Heinson, {\it et al}., Phys. Rev. {\bf D51} (1995) 985.

\bibitem{ygr}Y. Grossman and Y. Nir. Phys. Lett. {\bf B398} (1997) 163.

\bibitem{KEK}T. Akagi, {\it et al}., Phys. Rev. {\bf D51} (1995) 2061.

\bibitem{gab}F. Gabbiani, hep-ph/9901262.

\bibitem{buc}G. Buchalla, A.J. Buras, hep-ph/9901288.

\bibitem{adl}S. Adler, {\it et al}., Phys. Rev. Lett. {\bf B79} (1997) 2204.

\bibitem{lib}L. Littenberg, Phys. Rev. {\bf D39} (1989) 3322.

\bibitem{kk1}G. Buchalla, A.J. Buras, Nucl. Phys. {\bf B400} (1993) 225;
{\bf B548} (1999) 309.

\bibitem{new1}T. Hattori, T. Hasuike, S. Wakaizumi, hep-ph/9804412.

\bibitem{des}G.W. Kilcup, Nucl. Phys. (Proc. Suppl.) {\bf B20} (1991 417;
S.R. Sharpe, {\it ibid.}, 429;
D. Pekurovsky and G.Kilcup, hep-lat/9709146.

\end{thebibliography}
\end{document}